\documentclass[twocolumn]{aastex63}

\usepackage{longtable}
\usepackage{graphicx}
\usepackage{txfonts}
\usepackage{verbatim}
\usepackage{color}
\usepackage{threeparttablex}
\submitjournal{ApJ}

\newcommand{\msun}{$M/M_{\odot}\,$}

\usepackage{natbib}
\defcitealias{DeSomma2020}{DS2020}
\defcitealias{Ripepi2019}{Ripepi2019}
\defcitealias{Bhardwaj2017}{Bhardwaj2017}
\received{2020 June 10}
\revised{2020 June 19}
\accepted{2020 June 29}
\submitjournal{ApJL}
\shorttitle{Predicted masses of Galactic Cepheids}
\shortauthors{Marconi et al.}

\graphicspath{{./}{figures/}}

\begin{document}

\title{Predicted masses of Galactic Cepheids in the Gaia Data Release 2}

\correspondingauthor{Giulia De Somma}
\email{giulia.desomma@inaf.it , gdesomma@na.infn.it}

\author{Marcella Marconi}
\affiliation{ INAF-Osservatorio astronomico di Capodimonte \\
Via Moiariello 16 \\
80131 Napoli, Italy}

\author{Giulia De Somma}
\affiliation{ INAF-Osservatorio astronomico di Capodimonte \\
Via Moiariello 16 \\
80131 Napoli, Italy}
\affiliation{Dipartimento di Fisica ”E. Pancini”, Universit\'a di Napoli ”Federico II”\\
Compl. Univ. di Monte S. Angelo, Edificio G, Via Cinthia\\
I-80126, Napoli, Italy}
\affiliation{INFN-Sez. di Napoli\\
Compl. Univ.di Monte S. Angelo, Edificio G, Via Cinthia \\
I-80126, Napoli, Italy}

\author{Vincenzo Ripepi}
\affiliation{ INAF-Osservatorio astronomico di Capodimonte \\
Via Moiariello 16 \\
80131 Napoli, Italy}

\author{Roberto Molinaro}
\affiliation{ INAF-Osservatorio astronomico di Capodimonte \\
Via Moiariello 16 \\
80131 Napoli, Italy}

\author{Ilaria Musella}
\affiliation{ INAF-Osservatorio astronomico di Capodimonte \\
Via Moiariello 16 \\
80131 Napoli, Italy}

\author{Silvio Leccia}
\affiliation{ INAF-Osservatorio astronomico di Capodimonte \\
Via Moiariello 16 \\
80131 Napoli, Italy}

\author{Maria Ida Moretti}
\affiliation{ INAF-Osservatorio astronomico di Capodimonte \\
Via Moiariello 16 \\
80131 Napoli, Italy}

\begin{abstract}
\noindent 
On the basis of recently computed nonlinear convective pulsation
models of Galactic Cepheids, spanning wide ranges of input stellar
parameters, we derive theoretical mass-dependent Period-Wesenheit
relations in the Gaia bands, namely $G$, $G_{BP}$ and $G_{RP}$, that
are found to be almost independent of the assumed efficiency of super-adiabatic convection. The application to a selected sub-sample of Gaia Data Release 2 Galactic Cepheids database allows us to derive mass-dependent estimates of their individual distances.
By imposing their match with the astrometric values inferred from Gaia, we are able to evaluate the individual mass of each pulsator.
The inferred mass distribution is peaked around 5.6$M_{\odot}$ and 5.4$M_{\odot}$ for the F and FO pulsators, respectively. 
If the estimated Gaia parallax offset $<\Delta\varpi>$=0.046 mas is applied to Gaia parallaxes before imposing their coincidence with the theoretical ones, the inferred mass distribution is found to shift towards lower masses, namely $\sim$5.2$M_{\odot}$ and 5.1$M_{\odot}$ for the F and FO pulsators, respectively.  
The comparison with independent evaluations of the stellar masses, for a subset of binary Cepheids in our sample, seems to support the predictive capability of current theoretical scenario.
By forcing the coincidence of our mass determinations with these literature values we derive an independent estimate of the mean offset to be applied to Gaia DR2 parallaxes, $<\Delta\varpi>$=0.053 $\pm$ 0.029 mas, slightly higher but in agreement within the errors with \citet{Riess2018} value.

\end{abstract}


\section{Introduction} \label{sec:intro}
Classical Cepheids (CC) are massive and intermediate-mass ($\sim$ 3 - 13
$M_{\odot}$) stars crossing the pulsation instability strip while evolving
along the central helium burning phase \citep[see e.g.][and references therein]{Anderson2016,Bono2000a,CWC93}.
Thanks to their characteristic Period-Luminosity (PL) and
Period-Luminosity-Color (PLC) relations, they are considered the most
important primary distance indicators within the Local Group, currently
adopted to calibrate secondary distance indicators and, in turn, to
evaluate the Hubble constant \citep[see e.g.][and references therein]{Freedman2001, Riess2011, Riess2018, Riess2019, Ripepi2019}.
From the physical point of view, the occurrence of PL and PLC relations relies on the existence of the period-mean density relation coupled with the Stefan-Boltzmann law and the Mass-Luminosity (ML) relation predicted by stellar evolution models for central helium burning massive and intermediate-mass stars \citep[see e.g.][]{BCCM99,Bono2000a,CWC93}.
This implies that any phenomenon affecting the CC ML relation also affects the coefficients of the resulting PL and PLC relations and, in turn, the associated distance scale.
Theoretical evaluations of Cepheid masses based on stellar evolution models depend on the assumed ML relation \citep[see e.g.][]{CassisiSalaris2011} that is affected by chemical composition  and physical ingredients such as opacity \citep[see e.g. the new study by][suggesting that opacity might be underestimated]{Bailey2015}, equation of state and nuclear cross sections as well as by macroscopic phenomena, such as core overshooting, mass loss and rotation. On the other hand, theoretical attempts to derive Cepheid masses from stellar pulsation \citep[see e.g.][and references therein]{Bono2001, Caputo2005,KW2006,Marconi2013a,Marconi2013b,Marconi2017,Ragosta2019} do provide systematically lower masses than evolutionary estimates unless the latter adopt a moderate efficiency of core overshooting in the previous hydrogen burning phase, and/or mass loss and/or rotation. All these effects make the ML relation brighter than for canonical no mass loss, no rotation and no overshooting models. We notice that such a moderately brighter ML relation also allows us to match dynamical stellar mass derivations for Cepheids in eclipsing binary systems \citep[see e.g.][]{Marconi2013a, Neilson2012, Pietrzynski2010, Pietrzynski2011, PradaMoroni2012}.
In a recent theoretical investigation of Galactic Classical Cepheid (GCC) properties \citep[see][hereafter DS2020]{DeSomma2020} based on nonlinear convective models \citep[see][and references therein, for the physical and numerical assumptions]{BMS99,Marconi2005}, we predicted the light curves and the mean magnitudes and colors of solar chemical composition GCC in the Gaia filters, $G$, $G_{BP}$ and $G_{RP}$, varying both the ML relation and the efficiency of super-adiabatic convection. The inferred Period-Wesenheit (PW) relations were applied to a sample of Gaia Data Release 2 (hereinafter DR2) to constrain their individual distances and parallaxes \citepalias[see e.g.][for details]{DeSomma2020}.  
The results of this procedure and the comparison of the obtained values with Gaia DR2 observed parallaxes \citep[see][]{Clem2019,Ripepi2019} were found to depend on the assumed ML relation.
In this paper we reverse the perspective and rely on Gaia DR2 \citep{Brown2018, prusti2016} parallaxes to constrain GCC individual masses through inversion of predicted mass-dependent PW relations, thus testing a tool that will be fully efficient when the final Gaia data release will be available.
The organization of the paper is the following. In Section 2, we derive the mass-dependent PW relations from the nonlinear convective models computed by \citetalias{DeSomma2020}. In Section 3, we present the selected Cepheid sample and the procedure to derive individual masses. In Section 4, we compare the obtained individual masses with independent results for Cepheids in binary systems in the literature. Finally, Section 5 includes a discussion of results with some future developments.

\section{The mass-dependent Period-Wesenheit relations}

The predicted intensity weighted mean magnitudes and colors in the Gaia filters, $<G>$, $<G_{BP}>$ and $<G_{BR}>$, provided by \citetalias{DeSomma2020} for the extensive grid of pulsation models computed at solar chemical composition, depend on the assumed ML relation. In particular, for each stellar mass, three luminosity levels are considered in that paper, corresponding to a canonical value (case A, neglecting core-overshooting, mass loss and rotation effects) based on evolutionary predictions by \citet{Bono2000a} and two additional non-canonical luminosity levels obtained by increasing the canonical luminosity by 0.2 dex (case B) and 0.4 dex (case C).
Considering this whole model set, for each combination of mass, luminosity and effective temperature we can provide the corresponding predicted period and  Wesenheit function\footnote{the Gaia filter Wesenheit function is defined as $$W=<G>-1.9(<G_{BP}>-<G_{BR}>)$$ following the prescriptions by \citet{Ripepi2019}.} and, in turn, derive the mass-dependent Period-Wesenheit (hereinafter PWM) relations for the fundamental (F) and first overtone (FO) models, on the same period range as for the observed GCC.
The coefficients of the predicted relations for both F and FO models are reported in Table \ref{wpm_f_fo} for the two assumptions on the efficiency of super-adiabatic convection, namely $\alpha=1.5$\footnote{$\alpha$=$~l/H_{P}$ where l is the length of the path covered by the convective elements and $H_{P}$ is the local pressure height scale.} and $\alpha=1.7$.
We notice that a variation in the $\alpha$ parameter does not significantly affect the coefficients of the PWM relations, in spite of significant effects on the amplitude and morphology of light curves \citepalias[see][for details]{Bhardwaj2017,DeSomma2020}. For this reason, in the following we only consider model predictions for $\alpha=1.5$.
Among other parameters involved in the time-dependent convective treatment \citet[see][for details]{BonoStellingwerf1994}, the eddy  viscosity coefficient $\nu_{eddy}$ is set independently of the mixing length, whereas the overshooting length scale is related to $\alpha$. A variation of $\nu_{eddy}$ is expected to produce similar effects on the light curve amplitudes, morphology and the instability strip width as the $\alpha$ changes, but not to significantly affect the derivation of the PWM relations.
Moreover, we notice that, for Cepheid samples at the same distance, such relations  allow us to constrain the stellar mass distribution, whereas in the case of available individual distances, as for the Gaia database, the absolute individual mass values are directly determined.
In Figure \ref{wpm_f_fo} we plot the derived F (green symbols) and FO (red symbols) model distribution in the $W-c\log{M}~vs~ \log{P}$ plane, over-imposed to the projection of the inferred PWM relations. These relations will be used in the following section to infer individual mass estimates for a sample of GCC with Gaia DR2 distances.

\begin{table*}
\caption{\label{wpm_f_fo} The coefficients of the PWM relation ($W=a+b \log P+c \log M/M_{\odot}$) predicted for the F and FO-mode GCC, varying the mixing length parameter. The last column represent the root-mean-square deviation ($\sigma$) coefficient.}
\centering
\begin{tabular}{cccccccc}
\hline\hline
$\alpha_{ml}$&a&b&c&$\sigma_{a}$&$\sigma_{b}$&$\sigma_{c}$&$\sigma$\\
\hline
F\\
\hline
1.5&-1.654&-2.419&-2.423&0.036&0.021&0.067&0.064\\
1.7&-1.686&-2.496&-2.285&0.040&0.026&0.082&0.058\\
\hline
FO\\
\hline
1.5&-2.162&-3.068&-1.819&0.023&0.020&0.044&0.013\\
1.7&-2.205&-3.093&-1.765&0.032&0.027&0.062&0.008\\
\hline\hline
\end{tabular}
\end{table*}

\begin{figure}
\centering
\includegraphics[width=0.5\textwidth]{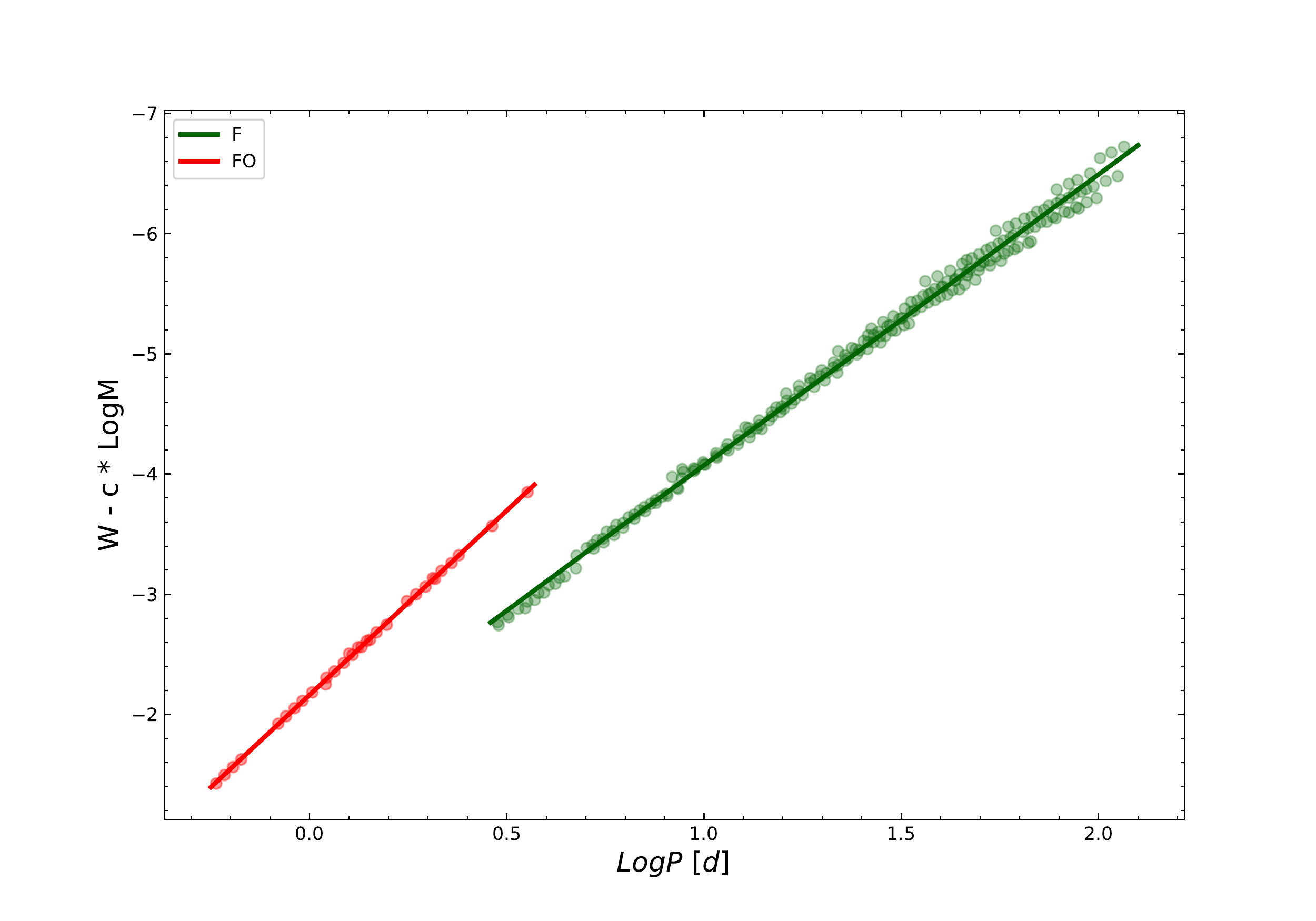}
\caption{Projection of the inferred PWM relations and F (green symbols) and FO (red symbols) model distribution in the $W-c\log{M} vs \log{P}$ plane.}
\label{fig:wpm_f_fo}
\end{figure}

\section{Application to Gaia DR2 Galactic Cepheids}

In this Section, we present a first test of the predictive capability of the derived PWM relations for the F and FO pulsators through their application to a subset of Gaia DR2 GCC \citepalias[see][]{DeSomma2020,Ripepi2019}.

\subsection{The selected sample}
The adopted sample of Gaia DR2 GCC is the one compiled by \citet{Ripepi2019} and used in \citetalias{DeSomma2020} to derive theoretical distances.
In the present work, in order to convert the observed Gaia parallaxes into distance moduli $\mu_{Gaia}$, then used to correct apparent Wesenheit magnitudes, we selected only Cepheids in the \citet{Ripepi2019} sample with a relative error on Gaia DR2 parallax lower than 10\% and positive mean parallax values. 
Table \ref{mass_F_FO_1.5}, from column 1 to 8 reports the Gaia source identification, the pulsation mode, the pulsation period, the mean apparent magnitudes in the Gaia filters, the measured parallax and the associated uncertainty of the selected GCC.
In the following we use these observed properties to constrain the individual stellar masses, through application of theoretical PWM relations.

\subsection{Derivation of individual Cepheid masses}

From the equation $$W_{oss}-\mu_{Gaia}=W_{teo}=a+b\log{P}+c\log{M/{M_{\odot}}}$$
where $W_{oss}$ is defined as $$W_{oss}=<G>-1.9(<G_{BP}>-<G_{BR}>)$$,
we are able to derive the stellar mass for each individual F and FO-mode pulsator.
The inferred stellar masses with the associated errors\footnote{The estimated errors take into account the uncertainty on the individual Gaia parallaxes, the intrinsic dispersion of the predicted PWM relations and the error on the estimated $W_{oss}$ considering a mean photometric error on the Gaia mean magnitudes of the order of 0.02 mag.}, for the F and FO-mode models, are reported in columns 9 and 10 of Table \ref{mass_F_FO_1.5}.

\begin{figure}
\centering
\includegraphics[width=0.4\textwidth]{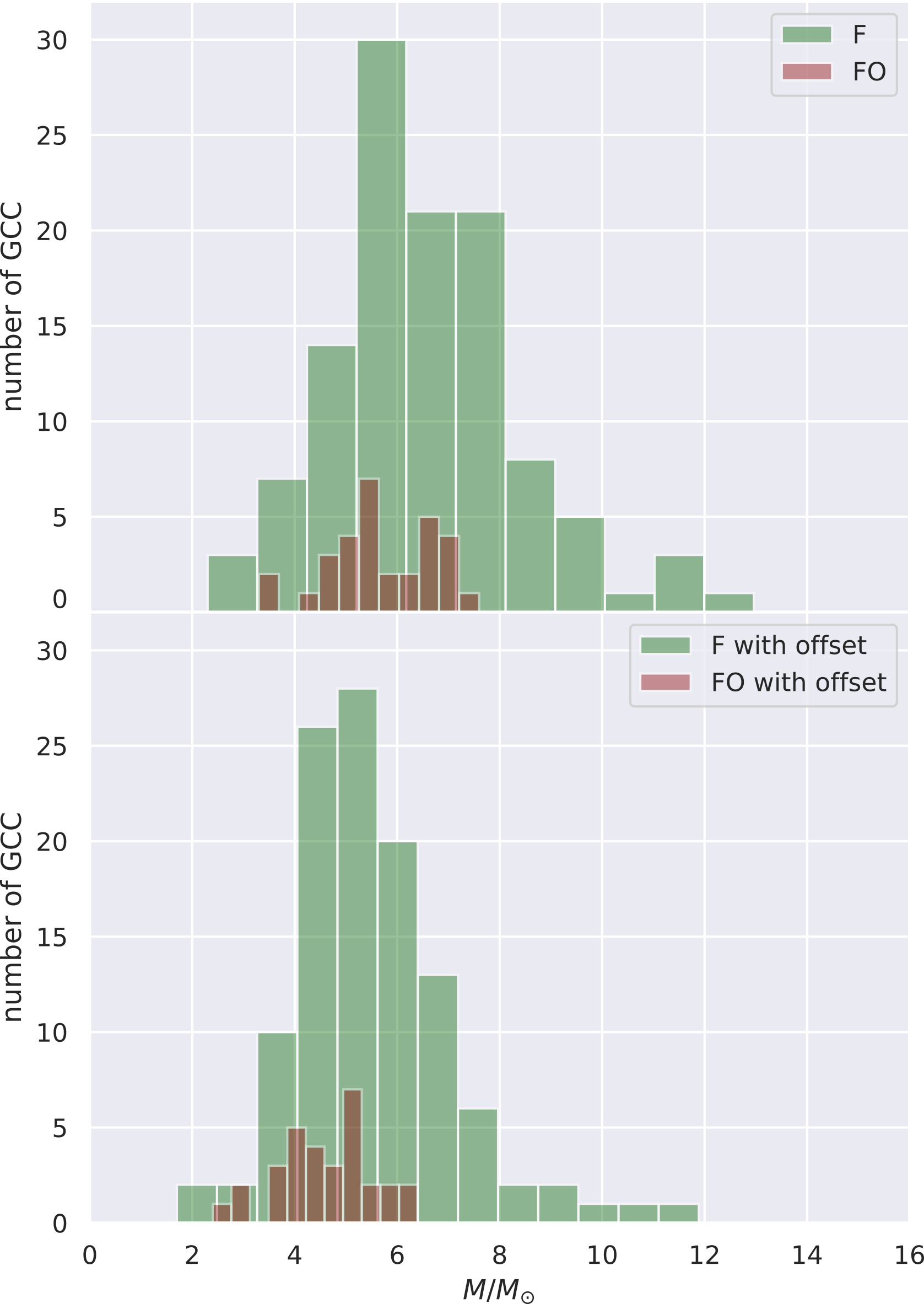}
\caption{\sl{Top panel}: The predicted mass distribution of the F (green bars) and FO-mode (red bars) pulsators. \sl{Bottom panel}: The same distribution as in the upper panel but obtained including the Gaia DR2 Cepheid parallax offset.}
\label{fig:hist_F_FO_tog}
\end{figure}

The upper panel of Figure \ref{fig:hist_F_FO_tog} shows the derived mass distribution histograms for the selected F (green bars) and FO (red bars) Gaia DR2 GCC.

We notice that the selected GCC sample is predicted to cover a relatively wide range of masses, peaked around 5.6$M_{\odot}$ and 5.4$M_{\odot}$ for the F and FO-mode pulsators, respectively.
Interestingly enough, if the error on the measured parallaxes decreased, as expected in the next Gaia Data Releases, we would obtain a corresponding improvement in the precision of our masses determination.
In particular, a precision on parallaxes of the order of 1\% would imply an error on the inferred stellar mass of the order of 2\% and 3\% in the case of the F and FO pulsators, respectively.

\begin{table*}
\caption{\label{mass_F_FO_1.5} The individual masses estimated from the theoretical PWM relations combined with Gaia DR2 parallaxes, for the F and FO-mode GCC in the selected sample. This table is available in its entirety in machine-readable form.}
\centering
\resizebox{\textwidth}{!}{%
\begin{tabular}{cccccccccccc}
\hline\hline
\textsl{Gaia} DR2 Source Id & Mode & P[d] & $G$[mag] & $G_{BP}$[mag] & $G_{RP}$[mag] & $\varpi$[mas] & $\sigma$$\varpi$[mas] & \msun & $\sigma$\msun & \msun{corr} &  $\sigma$\msun{corr}\\
\hline
(1)&(2)&(3)&(4)&(5)&(6)&(7)&(8)&(9)&(10)&(11)\\
\hline
1857884212378132096 & F & 4.43546 & 5.46 & 5.77 & 5.07 & 1.674 & 0.089 & 4.2 & 0.5 & 4.0 & 0.5 \\
4066429066901946368 & F & 5.05787 & 6.82 & 7.37 & 6.23 & 1.119 & 0.053 & 5.2 & 0.6 & 4.8 & 0.5 \\
5235910694044165760 & F & 3.08613 & 8.70 & 9.22 & 8.06 & 0.681 & 0.032 & 4.1 & 0.5 & 3.6 & 0.4 \\
... & ... & ... & ... & ... & ... & ... & ... & ... & ... \\
5351436724362450304 & FO & 1.11936 & 11.09 & 11.62 & 10.41 & 0.389 & 0.030 & 3.3 & 0.7 & 2.4 & 0.5 \\
2164475809937299584 & FO & 1.76585 & 10.18 & 10.74 & 9.49 & 0.343 & 0.027 & 7.6 & 1.7 & 5.4 & 1.2 \\
5245796334347122944 & FO & 2.06344 & 8.09 & 8.53 & 7.54 & 0.858 & 0.026 & 3.5 & 0.4 & 3.0 & 0.3 \\
... & ... & ... & ... & ... & ... & ... & ... & ... & ... \\
\hline\hline
\end{tabular}}
\end{table*}

\subsection{The effect of the Gaia parallax offset}

\begin{figure}
\centering
\includegraphics[width=0.4\textwidth]{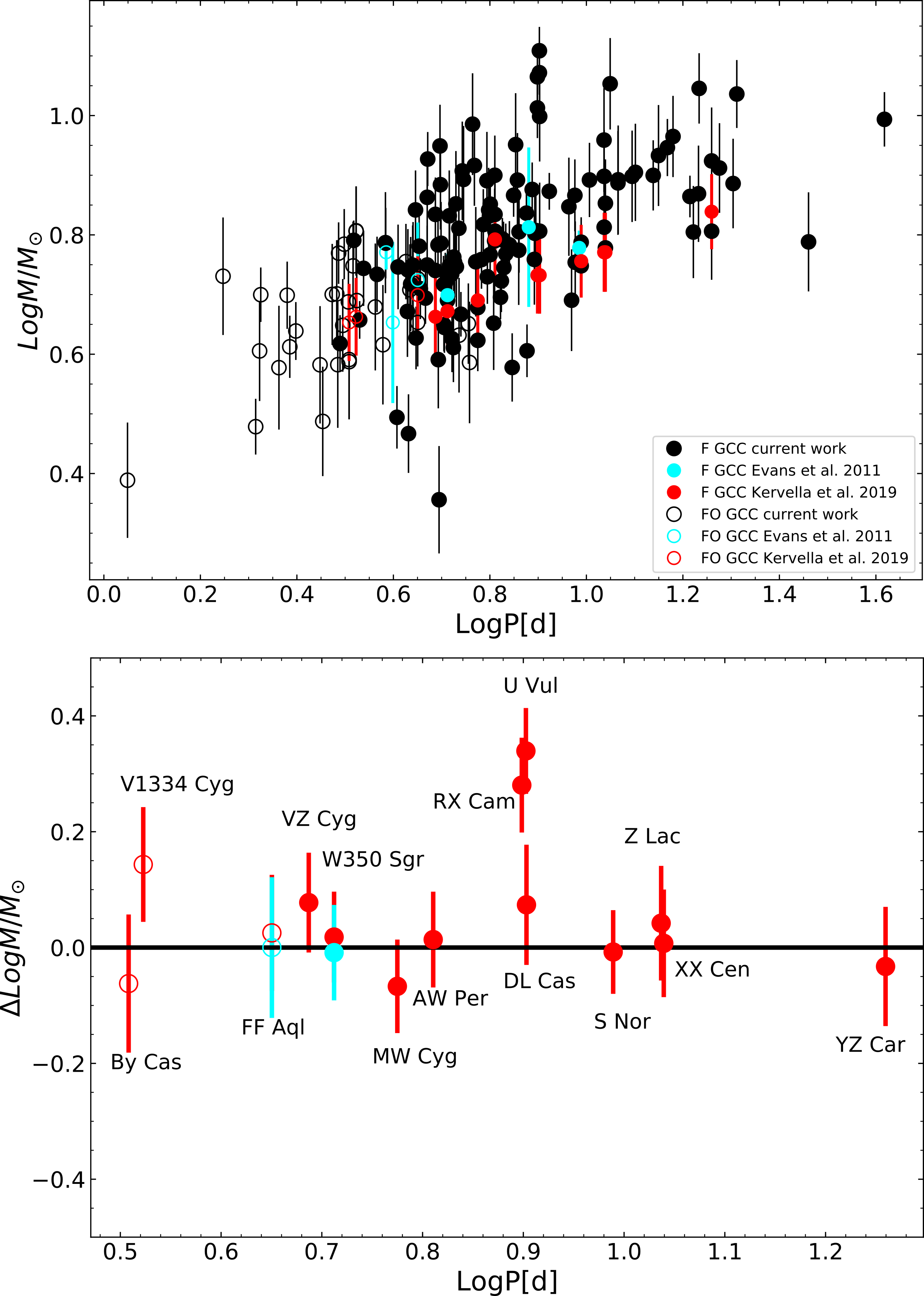}
\caption{\sl{Top panel}:The predicted mass distribution of the F (filled circles) and FO (open circles) pulsators as a function of the pulsation period. \sl{Bottom panel:} The difference between our results and the ones by \citet[][red symbols]{Kervella2019} and \citet[][cyan symbols]{Evans2011} for the Cepheids in common with the two data sets.}
\label{fig:comparison}
\end{figure}

To take into account the Gaia DR2 Cepheid parallax offset corresponding to $<\Delta\varpi>$=0.046$\pm$ 0.013 mas and derived by \citet{Riess2018} through the comparison with HST space scan astrometric determinations \citep[see][for details]{Riess2018}, we performed again our mass derivation procedure for the F and FO GCC, by increasing the parallax values reported in Table \ref{mass_F_FO_1.5} by $<\Delta\varpi>$=0.046 mas. The new estimated masses and the relative errors for the F and FO-mode pulsators are reported in the last two columns of the aforementioned Table. The obtained results are shown in the bottom panel of Figure \ref{fig:hist_F_FO_tog}.
We notice that the parallax offset effect moves the peak of the distribution to lower masses, around 5.2 $M_{\odot}$ and 5.1 $M_{\odot}$, for the F and FO mode, respectively.
This occurrence is expected on the basis of the coefficients of the PWM relations.
Indeed, an increase of the parallax implies a decrease in the distance modulus and, in turn, a fainter Wesenheit function,that at a fixed period, implies a lower mass. For the same reason, if the applied offset were $<\Delta\varpi>~=0.046+0.013=0.059$ mas, the inferred masses would be on average smaller than the literature ones, while if an offset $<\Delta\varpi>~=0.046-0.013=0.033$ mas were assumed, the inferred masses would become more discrepant with the literature ones with respect to Figure \ref{fig:comparison}

\subsection{Comparison with the literature}

In the upper panel of Figure \ref{fig:comparison} we show the behaviour of the theoretical masses derived with the PWM relations including the DR2 parallax offset, as a function of the pulsation period, for the F (filled circles) and FO-mode (open circles) GCC, compared with the position of the Cepheids in binary systems  for which independent mass estimates are available in \citet[][red symbols]{Kervella2019} and \citet[][and references therein, cyan symbols]{Evans2011}.
The general trend predicted by our theoretical scenario is in good agreement with the data. To better quantify this agreement, in the lower panel we show the difference between our results and the ones by \citet[][red symbols]{Kervella2019} and \citet[][cyan symbols]{Evans2011} for the Cepheids in common with the two data sets.
This plot confirms that we find a good agreement for most of the stars with the exception of RX Cam and U Vul for the F-mode. We notice that these two stars also deviate from more than 1 $\sigma$ from the empirical PW relation derived by \citet{Ripepi2019}. For the FO V1334 Cyg our estimate with the assumed offset and the result by \citet{Kervella2019} are quite different but still consistent within the errors. We also verified that a worse agreement with literature mass values is obtained when no offset is applied to Gaia parallaxes.

\section{Discussion}

The results shown in the previous section suggest that a general good agreement can be found between our mass determinations based on Gaia DR2 parallaxes combined with new derived theoretical PWM relations and independent mass values obtained for Cepheids in binary systems in the literature. 
This occurrence supports the accuracy of current theoretical scenario and at the same time paves the way to future applications.
In particular,  we plan to apply the same theoretical tool to the next more accurate Gaia Data Releases in order to reduce the error on mass determinations at the level of few \% with relevant implications for our knowledge of both the present mass function and the ML relation of intermediate-massive He-burning stars in the Milky Way. Moreover, by extending the PWM relation to other bands (including LSST Vera Rubin filters) and chemical compositions, we will be able to: i) infer the mass distributions of Cepheid samples in the Local Group for which accurate distances, e.g. LSST astrometric distances, will become available; ii) to constrain the coefficients of chemical abundances in theoretical Cepheid ML relations; iii) to predict the implications for the dependence of Cepheid properties and distance scale on the chemical composition. We notice that the PWM relation is expected to depend on metallicity because as the metallicity decreases, the theoretical quantity Mag-$1.9$*color is expected to get slightly fainter than in the solar case, according to previous results \citep[see e.g. figure 9 in][]{Caputo2000}.
Moreover, preliminary tests in the optical bands, based on the quoted previously computed models, suggest that the mass dependence of the PWM relation is reduced in lower metallicity model sets, with the effect of predicting systematically higher masses at a fixed distance and period.
On the other hand, by forcing the coincidence, within the errors, of our "uncorrected" mass evaluations  as reported in columns 9 and 10 of Table \ref{mass_F_FO_1.5}, with the literature determinations by Kervella et al. and Evans et al., reported in Figure \ref{fig:comparison}, we can derive an independent estimate of the offset that should be applied to Gaia DR2 parallaxes.
In particular, by excluding RX Cam,  U Vul and DL Cas, as well as V1334 Cyg, that deviate by more than 1 $\sigma$ from the PW relations by \citet{Ripepi2019},  we obtain a  mean offset $<\Delta\varpi>$= $0.053\pm0.029$ mas, where the uncertainty is the standard error of the mean. This result is  slightly higher than, but consistent within the errors, with the value obtained by Riess et al.

\acknowledgments
We thank the Referee for her/his pertinent and useful comments that significantly improved the content of the letter.
We acknowledge Istituto Nazionale di Fisica Nucleare (INFN), Naples section, specific initiative QGSKY. 
This work has made use of data from the European
Space Agency (ESA) mission \textsl{Gaia} (https://www.cosmos.esa.int/gaia),
processed by the \textsl{Gaia} Data Processing and Analysis Consortium (DPAC, https:
//www.cosmos.esa.int/web/gaia/dpac/consortium). Funding for the
DPAC has been provided by national institutions, in particular the institutions
participating in the Gaia Multilateral Agreement. In particular, the Italian participation
in DPAC has been supported by Istituto Nazionale di Astrofisica (INAF)
and the Agenzia Spaziale Italiana (ASI) through grants I/037/08/0, I/058/10/0, 2014-025-R.0, 2014-025-R.1.2015 and 2018-24-HH.0 to INAF (PI M.G. Lattanzi). We acknowledge partial financial support from 'Progetto Premiale' MIUR MITIC (PI B. Garilli) and the INAF Main Stream SSH program, 1.05.01.86.28.
This work has made use of the VizieR database, operated at CDS, Strasbourg, France.


\bibliography{mar}{}

\bibliographystyle{aasjournal}

\end{document}